\documentclass[3p,twocolumn]{elsarticle}

\usepackage{lineno,hyperref}
\usepackage{amsmath}
\usepackage{multirow}
\usepackage{booktabs}

\newcommand{\expect}[1]{\left<#1\right>}
\newcommand{\nuc}[2]{$^{#1}$#2}

\newcommand{\diff}{\text{d}}

\modulolinenumbers[5]

\journal{Physics Letters B}

\begin{document}

\begin{frontmatter}

\title{Compression-mode resonances in the calcium isotopes and implications for the asymmetry term in nuclear incompressibility}

\author[ND]{K. B. Howard}
\author[ND]{U. Garg \corref{cor1}}
\ead{garg@nd.edu}
\cortext[cor1]{Corresponding author}

\author[CYRIC]{M. Itoh}
\author[Konan]{H. Akimune}
\author[SMU,GSI,JUSTUS]{S. Bagchi}
\author[Kyoto]{T. Doi}
\author[Kyoto]{Y. Fujikawa}
\author[RCNP]{M. Fujiwara}
\author[Kyoto,RCNP]{T. Furuno}
\author[GSI,KVI]{M. N. Harakeh}
\author[Kyoto]{Y. Hijikata}
\author[Kyoto]{K. Inaba}
\author[CYRIC]{S. Ishida}
\author[KVI]{N. Kalantar-Nayestanaki}
\author[Osaka]{T. Kawabata}
\author[Konan]{S. Kawashima}
\author[Konan]{K. Kitamura}
\author[RCNP]{N. Kobayashi}
\author[CYRIC]{Y. Matsuda}
\author[CYRIC]{A. Nakagawa}
\author[RCNP]{S. Nakamura}
\author[Konan,CYRIC]{K. Nosaka}
\author[Kyoto]{S. Okamoto}
\author[CNS]{S. Ota}
\author[ND]{S. Weyhmiller}
\author[RCNP]{Z. Yang}

\address[ND]{Department of Physics, University of Notre Dame, Notre Dame, Indiana 46556, USA}
\address[CYRIC]{Cyclotron and Radioisotope Center, Tohoku University, Sendai 980-8578, Japan}
\address[Konan]{Department of Physics, Konan University, Hyogo 658-8501, Japan}
\address[SMU]{Astronomy and Physics Department, Saint Mary’s University, Halifax, NS B3H 3C3, Canada}
\address[GSI]{GSI Helmholtzzentrum f\"ur Schwerionenforschung GmbH, D-64291 Darmstadt, Germany}
\address[JUSTUS]{Justus-Liebig University, 35392 Giessen, Germany}
\address[Kyoto]{Department of Physics, Kyoto University, Kitashirakawa Oiwake, Sakyo, Kyoto 606-8502, Japan}
\address[RCNP]{Research Center for Nuclear Physics, Osaka University, Osaka 567-0047, Japan}
\address[KVI]{KVI-CART, University of Groningen, 9747 AA Groningen, The Netherlands}
\address[Osaka]{Department of Physics, Osaka University,
 Toyonaka, Osaka 540-0043, Japan}
\address[CNS]{Center for Nuclear Study, The University of Tokyo, Wako, Saitama 351-0198, Japan}

\begin{abstract}

Recent data on isoscalar giant monopole resonance (ISGMR) in the calcium isotopes \nuc{40,44,48}{Ca} have suggested that $K_\tau$, the asymmetry term in the nuclear incompressibility, has a positive value. A value of $K_\tau > 0$ is entirely incompatible with present theoretical frameworks and, if correct, would have far-reaching implications on our understanding of myriad nuclear and astrophysical phenomena. This paper presents results of an independent ISGMR measurement with the \nuc{40,42,44,48}{Ca}($\alpha,\alpha^\prime$) reaction at $E_\alpha = 386$ MeV. These results conclusively discount the possibility of a positive value for $K_\tau$, and are consistent with the previously-obtained values for this quantity.

\end{abstract}


\end{frontmatter}
The isoscalar giant monopole resonance (ISGMR) has been well-established as the most direct means by which one can constrain the incompressibility of nuclear matter.
The incompressibility of a nucleus is extracted from the resonance energy, $E_\text{ISGMR}$, such that:
\begin{align}
  E_\text{ISGMR} = \hbar \sqrt{\frac{K_A}{m \expect{r_0^2}}} \label{E_GMR},
\end{align}
where $K_A$ is the incompressibility of the nucleus of mass $A$ undergoing the excitation, $m$ is the free-nucleon mass, and $\expect{r_0^2}$ is the mean-square radius of the ground-state density. Within different model frameworks, the value of $E_\text{ISGMR}$ is associated with different moment ratios of the ISGMR strength distribution \cite{stringari_sum_rules,harakeh_book}, the extraction of which is the primary goal of many experiments venturing to constrain the incompressibility, $K_\infty$, and, thence, the Equation of State (EoS) of infinite nuclear matter \cite{garg_colo_review}.

In a system with a proton-neutron imbalance, the EoS depends additionally on the asymmetry parameter $\eta = (N-Z)/A$, and the symmetry energy, $S(\rho)$. In the same way that the ISGMR provides a direct measurement of $K_\infty$, the curvature of the EoS of symmetric nuclear matter, the trend of measurements of nuclei with varying values of $\eta$ yields a direct constraint of the curvature of $S(\rho)$. For a more complete discussion of the means by which properties of the giant resonances provide constraints on the EoS of such asymmetric nuclear matter, we refer the reader to Refs. \cite{garg_colo_review,colo_garg_sagawa_symmetry_energy_review}.

The microscopic formalism for extracting properties of infinite nuclear matter from measurements on finite nuclei is detailed in Ref. \cite{blaizot}. The macroscopic leptodermous expansion of $K_A$ in terms of properties of infinite nuclear matter gives:

\begin{align}
  K_A \approx K_\infty + K_\text{surf} A^{-1/3} + K_\tau \eta^2 + K_\text{Coul} \frac{Z^2}{A^{4/3}}. \label{lepto}
\end{align}

Equation \eqref{lepto} is useful in determining the value of $K_\tau$ for finite nuclei, owing in part to the isolated dependence on $\eta$ within the expression, as well as the fairly minimal changes in the surface term within an isotopic chain. The general prescription for the same is detailed in Refs. \cite{Li_PRL,Li_PRC}, and involves quadratically fitting the dependence of $K_A - K_\text{Coul} Z^2/A^{4/3}$ on $\eta$ with a model function of the form $K_\tau \eta^2 + c$, with $c$ being a constant. The $K_\tau$ values so extracted are consistent with one another:  $K_\tau = - 550 \pm 100$ MeV and $K_\tau = - 555 \pm 75$ MeV, for the even-$A$ \nuc{112-124}{Sn} and \nuc{106-116}{Cd} isotopes, respectively  \cite{Li_PRL,Li_PRC,Patel_PLB_Cd}.

The corresponding definition of $K_\tau^\infty$ in terms of properties of the EoS for infinite nuclear matter is\footnote{It should be noted that $K_\tau^\infty$ is \emph{not} equal to the value of $K_\tau$ extracted from finite nuclei utilizing the methodology of Eq. \eqref{lepto}, just as $K_\infty \neq K_A$. However, through the same self-consistent mechanisms by which measurements of $K_A$ serve to constrain $K_\infty$ as described by Blaizot \cite{blaizot}, determining values of $K_\tau$ from finite nuclei can constrain the EoS for asymmetric infinite nuclear matter.} \cite{Piekarewicz_centelles}:

\begin{align}
  K_\tau^\infty = K_\text{sym} - 6 L - \frac{Q_\infty}{K_\infty}L
  \label{KT_defined}
\end{align}
within which $L$ and $K_\text{sym}$ are, respectively, the slope and curvature of $S(\rho)$, and $Q_\infty/K_\infty$ is the skewness parameter for the EoS of symmetric nuclear matter. The implications of this are that experimental constraints on $K_\tau$ arising from measurements of $K_A$ on finite nuclei are helpful in determining the density dependence of the symmetry energy; this argument is predicated on the smoothness with which the values of $K_A$ vary across the nuclear chart.  Indeed, as has been argued in Ref. \cite{KBH_EPJA}, any nuclear structure effects on properties of the giant resonances which arise in a narrow region of the chart of nuclides would dramatically affect our understanding of the collective model upon which has rested the understanding of these resonances.

In light of all this, the results reported recently for \nuc{40,44,48}{Ca} by the TAMU group \cite{TAMU_40Ca,TAMU_44Ca,TAMU_48Ca} were very surprising: the moment ratios  for the ISGMR and, therefore, the $K_A$ values for \nuc{40,44,48}{Ca} \emph{increased} with increasing mass number. The most immediate consequence of this, considering Eq. \eqref{lepto}, is that $K_\tau$ is a \emph{positive} quantity, and it was shown in Ref. \cite{TAMU_44Ca} that a large positive value of $K_\tau$ models the data well. In a test of hundreds of energy-density functionals currently in use in the literature, the values of $K_\tau$ extracted were consistently between $-800$ MeV $ \leq K_\tau \leq -100$ MeV \cite{sagawa_private}. Examination of Eq. \eqref{KT_defined} also directly suggests that the symmetry energy would need to be extremely soft in order to accomodate $K_\tau > 0$ \cite{piekarewicz_private}. Moreover, the hydrodynamical model predicts $E_\text{ISGMR} \sim A^{-1/3}$, while the results of Refs. \cite{TAMU_40Ca,TAMU_44Ca,TAMU_48Ca} indicated exactly the opposite: the ISGMR energies increasing with mass number over the isotopic chain.

These results clearly demanded an independent verification before significant theoretical efforts were expended in understanding, and explaining, this unusual and unexpected phenomenon. This Letter presents the results of such an experiment; it is found that the ISGMRs in the measured calcium isotopes ($A=40$, $42$, $44$, $48$) follow the ``normal'' pattern of $E_\text{ISGMR}$ decreasing with increasing mass, {\em ruling out} a positive value for $K_\tau$.

The measurements were carried out at the Research Center for Nuclear Physics (RCNP) at Osaka University, utilizing a beam of 386-MeV $\alpha$-particles which was, for all practical purposes, ``halo-free''. This beam impinged onto enriched \nuc{40,42,44,48}{Ca} target foils of areal densities $\sim$1--3  mg/cm$^2$. The scattered $\alpha$-particles were momentum analyzed in the high-resolution magnetic spectrometer, Grand Raiden \cite{Fujiwara_Grand_Raiden}. The focal-plane detection system was comprised of a pair of vertical and horizontal position-sensitive multiwire drift chambers in coincidence with plastic scintillators which provided the particle identification signal \cite{tamii_grand_raiden, tamii_DAQ}. A recent and comprehensive description of the procedure employed in the offline data analysis has been presented in Ref. \cite{KBH_EPJA}. Here, we briefly revisit the most salient points: in addition to the lateral dispersion of the spectrometer allowing for the scattered particles to be distributed across the horizontal focal plane according to their momentum, the unique optical properties of Grand Raiden allow for $\alpha$-particles whose momentum transfers occurred only at the target to be coherently focused onto the median of the vertical focal plane distribution. On the other hand, particles that undergo scattering processes before or after the target are, correspondingly, over- or under-focused along the vertical axis. The latter events constitute the instrumental background, which in the present methodology can be eliminated from the inelastic spectra prior to further analysis.  This removes any, and all, ambiguities associated with modeling the instrumental background in the subsequent extraction of the ISGMR strength distributions. In contrast, the measurements reported in Refs. \cite{TAMU_40Ca,TAMU_44Ca,TAMU_48Ca} employed a phenomenological modeling of the instrumental background, thus introducing an additional uncertainty in the analysis \cite{youngblood_24Mg_continuum_study}; this process was suggested as the most likely reason behind differences in the extracted ISGMR strength distributions noted recently for the $A\sim90$ nuclei \cite{KBH_EPJA,youngblood_A90_unexpected,gupta_A90_PLB,gupta_A90_PRC}.

\begin{figure}[b!]
  \centering
  \includegraphics[width=\linewidth]{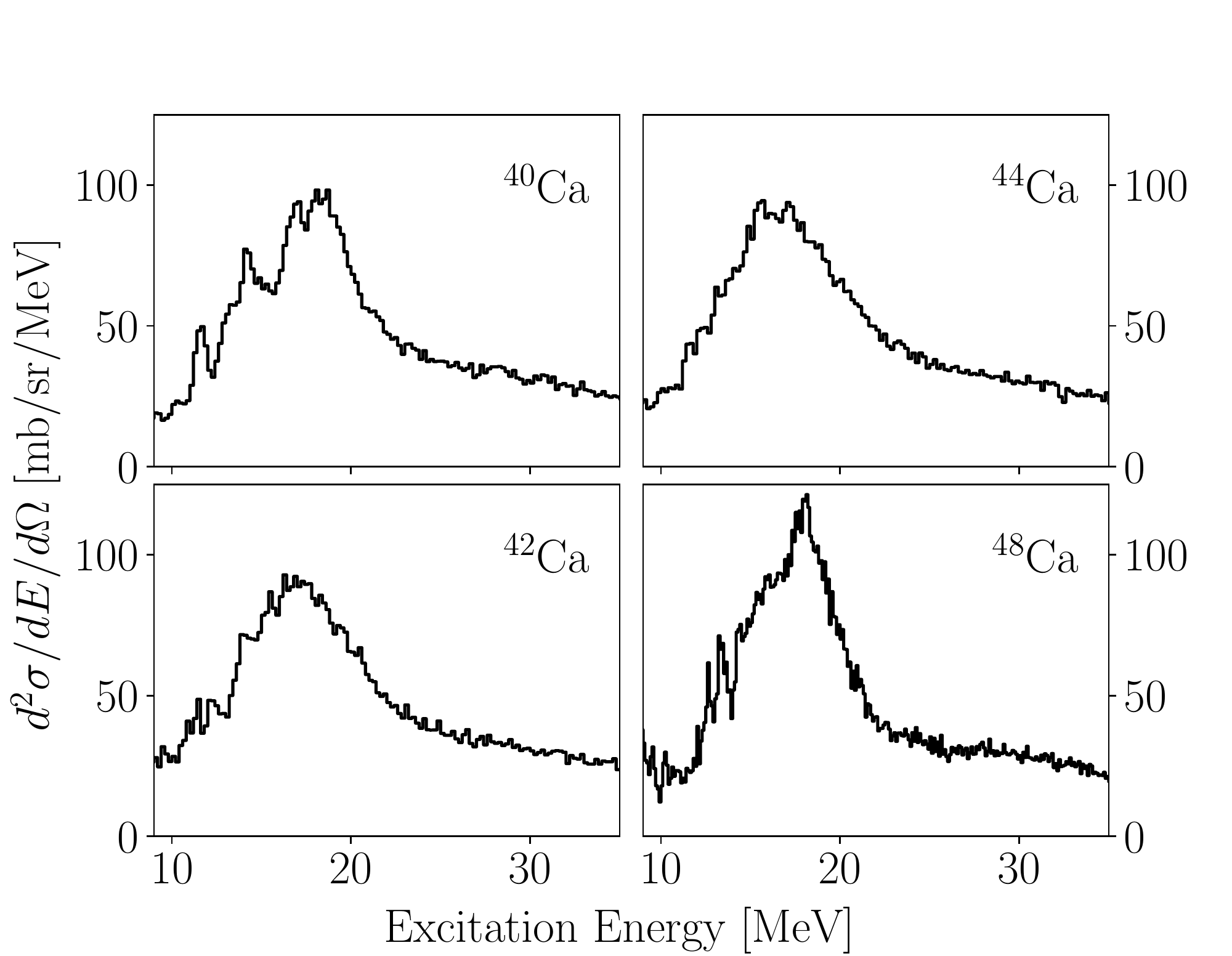}
  \caption{Measured double-differential cross-section spectra from \nuc{40,42,44,48}{Ca}($\alpha,\alpha^\prime)$ at $\theta_\text{Lab}=0.7^\circ$, after particle identification and subtraction of the instrumental background.}
  \label{calcium_spec}
\end{figure}

Inelastic scattering data were obtained over a broad angular range $0^\circ \leq \theta_\text{Lab} \leq 10^\circ$, and the acceptance of the spectrometer along the lateral dispersive plane ranged from approximately $10 \leq E_x \leq 35$ MeV. For each angular setting of the spectrometer, a precise multi-point energy calibration was acquired via the analysis of \nuc{24}{Mg}($\alpha,\alpha^\prime$) spectra. The energy loss through the target foils was accounted for within the model framework of SRIM \cite{SRIM}. The inelastic angular distributions were extracted in 200-keV-wide bins for \nuc{40,42,44}{Ca}; in order to achieve comparable statistical uncertainties, a wider 1-MeV-wide bin size was used for the analysis of \nuc{48}{Ca}. The ``0$^{\circ}$'' spectra, where the ISGMR cross sections are maximal, are presented in Fig. \ref{calcium_spec}.

To extract the strength distributions of the giant resonances, it is necessary to have an optical model parameter (OMP)-set to be employed in the Distorted-Wave Born Approximation (DWBA) calculations. To adequately constrain the OMP, elastic scattering data were measured for \nuc{42,44,48}{Ca} over a broader angular range ($5^\circ-25^\circ$). The nuclear reactions code \texttt{PTOLEMY} was utilized for a $\chi^2$ minimization of DWBA results from a single-folding, density dependent, hybrid optical model potential \cite{khoa_satchler_single_folding} relative to this data with the empirical density distributions reported in \cite{fricke_ADNDT}. The extracted OMPs are presented in Table~\ref{OMP_table}; further details of this procedure have been provided elsewhere \cite{KBH_EPJA}. Because of unavailability of elastic scattering data on  $^{40}$Ca, OMPs extracted for \nuc{42}{Ca} were employed for that nucleus. The use of OMPs from a nearby nucleus has been shown to have negligible effect in the results of the giant resonance strength extraction \cite{Li_PRC}, which is further evidenced by the minimal variation in the the OMPs themselves as seen in Table~\ref{OMP_table} for \nuc{42}{Ca} and \nuc{44}{Ca}. We further note that Refs. \cite{TAMU_44Ca} and \cite{TAMU_48Ca} has also employed the same OMP-set in the analyses of the \nuc{44}{Ca} and \nuc{48}{Ca} giant resonance data which originally motivated this work.

\begin{table}[t!]
\centering
\begin{tabular}{@{}lcccc@{}}
\toprule
                       & $V_\text{vol}$ & $W_\text{vol}$ & $R_I$ & $a_I$ \\
\multicolumn{1}{c}{} & [MeV]          & [MeV]          & [fm]  & [fm]  \\ \midrule
 \nuc{40}{Ca}         & 37.4           & 31.6           & 4.47  & 0.990 \\
\nuc{42}{Ca}         & 37.4           & 31.6           & 4.53  & 0.990 \\
\nuc{44}{Ca}         & 37.4           & 31.1           & 4.64  & 0.990 \\
\nuc{48}{Ca}         & 41.2           & 32.7           & 4.82  & 0.939 \\ \bottomrule
\end{tabular}
\caption[Optical model parameters extracted from fits to elastic scattering angular distributions and used for the DWBA input to the multipole-decomposition analyses.]{Optical-model parameters extracted from fits to elastic scattering angular distributions for \nuc{42,44,48}{Ca}. Definitions of the parameters are provided in Ref. \cite{KBH_EPJA}}
%which were later used for calculating the DWBA input to the multipole-decomposition analyses.}
\label{OMP_table}
\end{table}

\begin{figure}[t!]
  \centering
  \includegraphics[width=\linewidth]{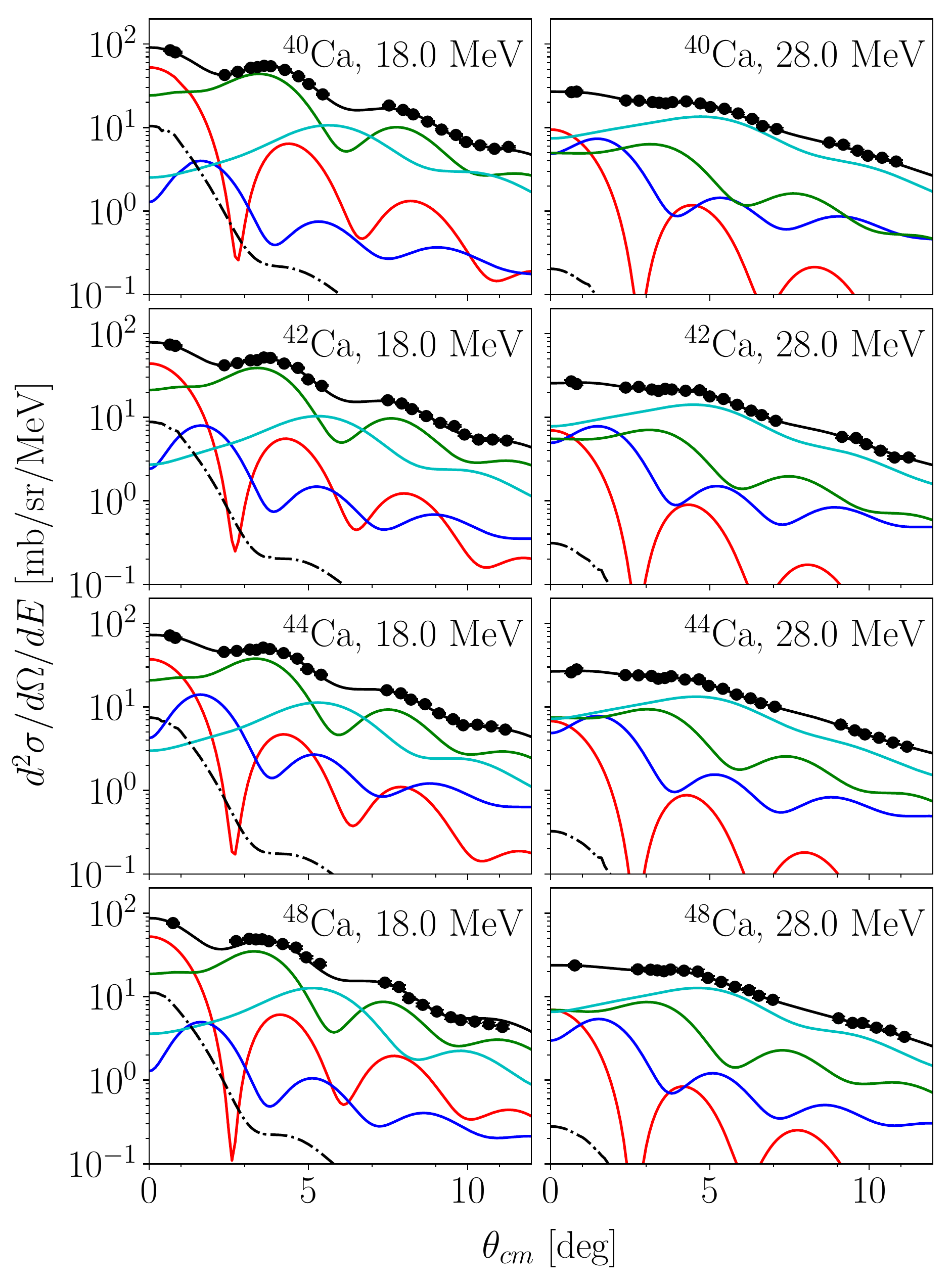}
  \caption{(Color online) Multipole-decomposition analyses for \nuc{40,42,44,48}{Ca} for excitation energy bins centered at 18 MeV (left panels) and 28 MeV (right panels). Shown are the total fits (solid black lines), as well as the fitted contributions from  the isoscalar monopole (red), dipole (blue), quadrupole (green), and higher multipole modes (cyan). Also shown is the contribution from the isovector giant dipole resonance (dot-dashed line), based on known photoneutron cross-section data and the Goldhaber-Teller model.}
  \label{calcium_MDA}
\end{figure}

The Multipole-Decomposition Analysis (MDA) of the inelastic spectra was carried out employing the now ``standard'' procedure, described, for example, in Refs. \cite{KBH_EPJA,gupta_A90_PRC,foreman_mackey,goodman_weare}. The experimental double-differential cross sections over the $E_x = 10-31$ MeV region were decomposed into a linear combination of the DWBA angular distributions for pure angular momentum transfers:
\begin{align}
  \frac{\diff^2 \sigma^\text{exp}(\theta_\text{c.m.},E_x)}{\diff \Omega \, \diff E_x} &= \sum_{\lambda} A_\lambda(E_x) \frac{\diff^2 \sigma_\lambda^\text{DWBA}(\theta_\text{c.m.},E_x)}{\diff \Omega \, \diff E_x}.
\end{align}
The $A_\lambda(E_x)$ coefficients correspond to the fraction of the energy-weighted sum rule (EWSR) for the multipolarity $\lambda$ exhausted within a particular energy bin \cite{harakeh_book}. DWBA cross sections for isoscalar modes were included in the MDA up to $\lambda_\text{max} = 8$, and the contribution from the isovector giant dipole resonance (IVGDR) was  accounted for using the Goldhaber-Teller model and the available photoneutron data for the calcium isotopes \cite{satchler_isospin,plujko_GDR}. Typical results of the MDA are presented in Fig. \ref{calcium_MDA}.

\begin{figure}[t!]
  \includegraphics[width=\linewidth]{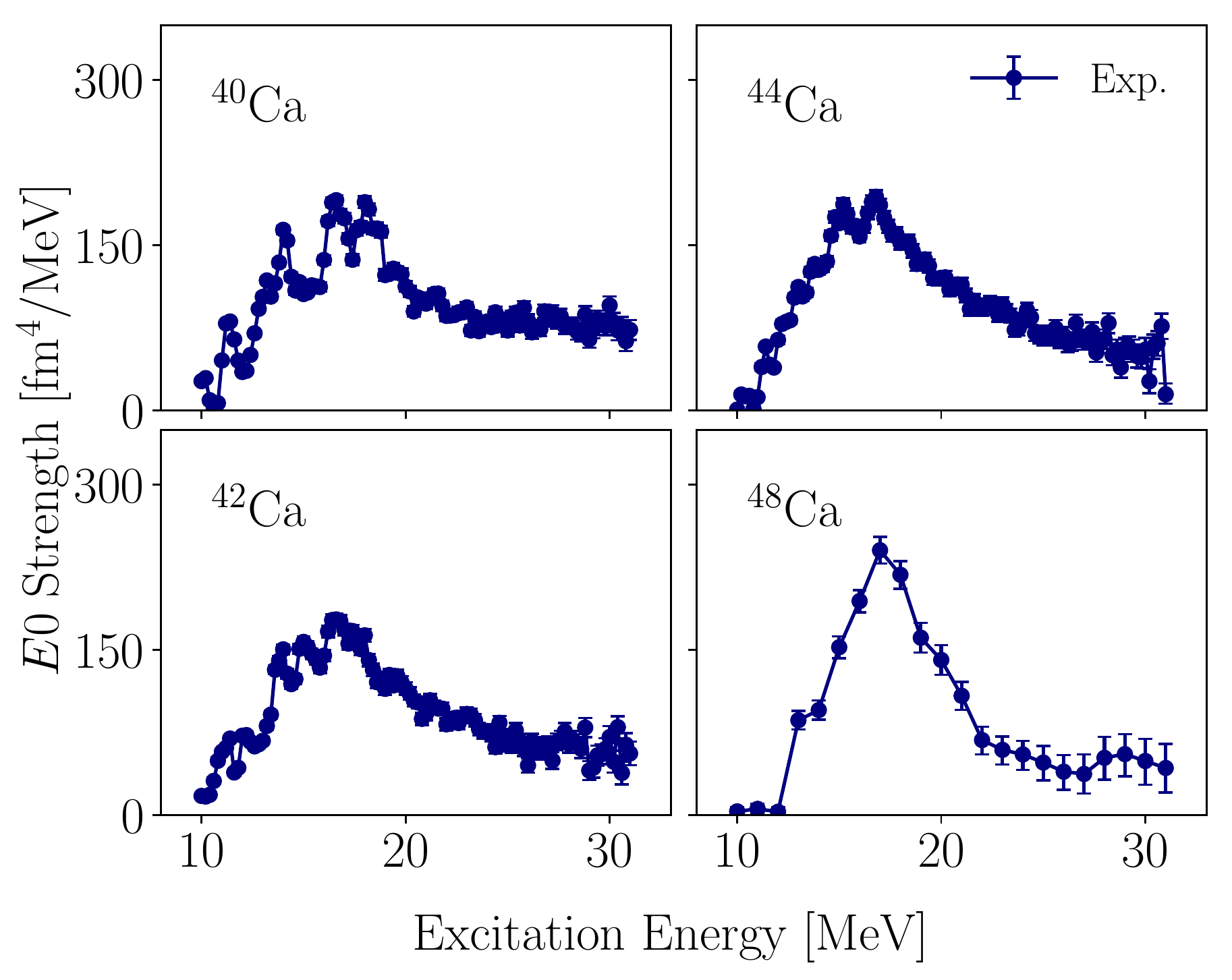}
  \caption{(Color online) Extracted isoscalar monopole strength distributions for \nuc{40,42,44,48}{Ca}.}
  \label{monopole strength}
\end{figure}

From the $A_\lambda$ coefficients, the strength distributions for the ISGMR were calculated using the corresponding EWSR relationships \cite{harakeh_book}. Shown in Fig. \ref{monopole strength} are the extracted ISGMR strength distributions for each of the calcium nuclei investigated in this work. From these extracted strength distributions, $S(E_x)$, the moments of the strength distribution were extracted in the usual way:
\begin{align}
  m_k &= \int S(E_x) E_x^k \, \diff E_x.
\end{align}

The moment ratios $\sqrt{m_1/m_{-1}}$, $m_1/m_0$, and $\sqrt{m_3/m_1}$ that are customarily used in characterizing the excitation energy of the ISGMR \cite{stringari_lipparini_sum_rules} are presented in Table \ref{moment_ratios_total_EWSRS}. The quoted uncertainties have been estimated using a Monte Carlo sampling from the probability distributions of the individual $A_\lambda(E_x)$ and constitute a $68\%$ confidence interval. The pattern of moment ratios observed in the calcium isotopic chain (decreasing with $A$, as expected from the $A^{-1/3}$ rule)  is contrary to that reported in Ref. \cite{TAMU_44Ca} {\em viz.} increase in the moment ratios with increasing $A$.

\begin{table}[h]
\centering
\begin{tabular}{@{}ccllcclclccll@{}}
\toprule
                    & \multicolumn{12}{c}{RCNP}                                                                                                                                                      \\ \cmidrule(l){2-13}
                    & \multicolumn{3}{c}{\nuc{40}{Ca}}         & \multicolumn{3}{c}{\nuc{42}{Ca}}         & \multicolumn{3}{c}{\nuc{44}{Ca}}           & \multicolumn{3}{c}{\nuc{48}{Ca}}            \\ \cmidrule(l){2-13}
\rule{0pt}{3ex} $m_1$ \%               & \multicolumn{3}{c}{     $102^{+3}_{-4}$  }              & \multicolumn{3}{c}{    $89^{+3}_{-3}$   }              & \multicolumn{3}{c}{  $88^{+4}_{-4}$ }                    & \multicolumn{3}{c}{       $78^{+4}_{-3}$ }                \\[10pt]
$\displaystyle \sqrt{\frac{m_1}{m_{-1}}}$ & \multicolumn{3}{c}{$19.5^{+0.1}_{-0.1}$} & \multicolumn{3}{c}{$19.0^{+0.1}_{-0.1}$} & \multicolumn{3}{c}{$18.9^{+0.1}_{-0.1}$}   & \multicolumn{3}{c}{$19.0^{+0.2}_{-0.2}$}    \\[10pt]
$\displaystyle {\frac{m_1}{m_{0}}}$           & \multicolumn{3}{c}{$20.2^{+0.1}_{-0.1}$} & \multicolumn{3}{c}{$19.7^{+0.1}_{-0.1}$} & \multicolumn{3}{c}{$19.5^{+0.1}_{-0.1}$}   & \multicolumn{3}{c}{$19.5^{+0.1}_{-0.1}$}    \\[10pt]
$\displaystyle \sqrt{\frac{m_3}{m_{1}}}$    & \multicolumn{3}{c}{$22.3^{+0.1}_{-0.1}$} & \multicolumn{3}{c}{$21.7^{+0.1}_{-0.1}$} & \multicolumn{3}{c}{$21.5^{+0.1}_{-0.1}$}   & \multicolumn{3}{c}{$21.3^{+0.3}_{-0.3}$}    \\ \cmidrule(l){2-13}
                    & \multicolumn{12}{c}{TAMU}                                                                                                                                                      \\ \cmidrule(l){2-13}
                    & \multicolumn{4}{c}{\nuc{40}{Ca}}                        & \multicolumn{4}{c}{\nuc{44}{Ca}}                        & \multicolumn{4}{c}{\nuc{48}{Ca}}                           \\ \cmidrule(l){2-13}
\rule{0pt}{3ex} $m_1$ \%               & \multicolumn{4}{c}{$97^{+11}_{-11}$ \hphantom{$0^{0}$}  }                                    & \multicolumn{4}{c}{$75^{+11}_{-11} $ }                                    & \multicolumn{4}{c}{$95^{+11}_{-15}$ \hphantom{$0^{0}$}}                                       \\[10pt]
$\displaystyle \sqrt{\frac{m_1}{m_{-1}}}$ & \multicolumn{4}{c}{$18.3^{+0.30}_{-0.30}$\hphantom{$0^{0}$}}  & \multicolumn{4}{c}{$18.73^{+0.29}_{-0.29}$}             & \multicolumn{4}{c}{$19.0^{+0.1}_{-0.1}$\hphantom{$0^{0}$}}      \\[10pt]
$\displaystyle {\frac{m_1}{m_{0}}}$           & \multicolumn{4}{c}{$19.2^{+0.40}_{-0.40}$\hphantom{$0^{0}$}}  & \multicolumn{4}{c}{$19.50^{+0.35}_{-0.33}$}             & \multicolumn{4}{c}{$19.9^{+0.2}_{-0.2}$\hphantom{$0^{0}$}}\\[10pt]
$\displaystyle \sqrt{\frac{m_3}{m_{1}}}$   & \multicolumn{4}{c}{$20.6^{+0.40}_{-0.40}$\hphantom{$0^{0}$}}  & \multicolumn{4}{c}{$21.78^{+0.84}_{-0.72}$}             & \multicolumn{4}{c}{$22.6^{+0.3}_{-0.3}$\hphantom{$0^{0}$}}       \\ \bottomrule
\end{tabular}
\caption{Percentages of the EWSR ($m_1$) for the ISGMR strength distributions, as well as the corresponding moment ratios (in MeV) calculated over the energy range $10-31$ MeV. The constrained-model ($\sqrt{m_1/m_{-1}}$), centroid ($m_1/m_0$), and scaling-model ($\sqrt{m_3/m_1}$) energies are presented. The corresponding quantities which were reported by the TAMU group are also shown for comparison; these were calculated over the energy range $9-40$ MeV \cite{TAMU_40Ca,TAMU_44Ca,TAMU_48Ca}. In all cases, the quoted uncertainties in the \%EWSR ($m_1$) are only statistical; there can be 15\%--20\% additional uncertainty from the DWBA calculations themselves (from the choice of the OMP, for example).}
\label{moment_ratios_total_EWSRS}
\end{table}

\begin{figure}[t]
  \includegraphics[width=\linewidth]{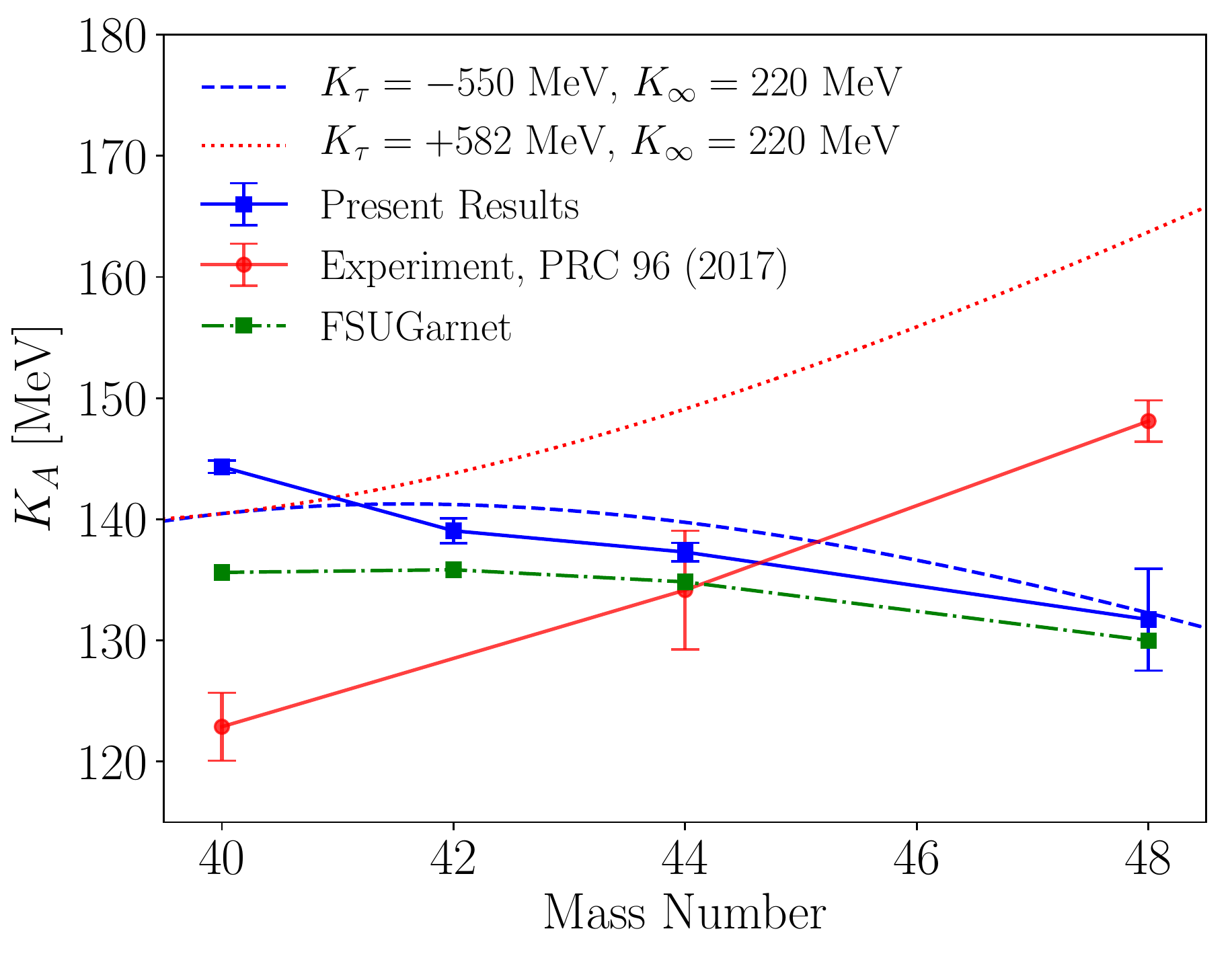}
  \caption{(Color online) The incompressibility, $K_A$, for the Ca isotopes investigated in this work (blue squares). These were calculated within the scaling model from the experimental data ($E_\text{ISGMR} = \sqrt{m_3/m_1}$, for consistency with the presentation of Ref. \cite{TAMU_44Ca}; see Table \ref{moment_ratios_total_EWSRS}). The expected trend for these values utilizing the previously documented central value for $K_\tau = -550$ MeV, and $K_\infty = 220$ MeV as input to Eq. \eqref{lepto} is presented (blue dashed line), along with the same calculation but with the value $K_\tau = +582$ MeV reported in Ref. \cite{TAMU_44Ca} (red dotted line). A fit to the data leads to a curve that is nearly identical to that shown above (blue dashed line) and leads to a value of $K_\tau = -510 \pm 115$ MeV. For comparison purposes, the data from Ref. \cite{TAMU_44Ca} are shown (red circles), as well as the $K_A$ values calculated from the ISGMR responses predicted by the relativistic FSUGarnet interaction (green squares) \cite{piekarewicz_private,piekarewicz_calcium_isotopes}. The solid lines through the data points are merely to guide the eye.}
  \label{RCNP KA}
\end{figure}

In addition to the moment ratios exhibiting the expected behavior over the isotopic chain, the demonstrated trend for the extracted finite incompressibilities, $K_A$, is even more illustrative (see Fig. \ref{RCNP KA}): The agreement of the extracted $K_A$ values with the behavior modeled by the leptodermous expansion of Eq. \eqref{lepto} using the accepted values for $K_\tau$ and $K_\infty$ is rather good, and stands in stark contrast to the results from Ref. \cite{TAMU_44Ca}. While the extracted $K_A$ for \nuc{44}{Ca} is consistent with that which was measured in Ref. \cite{TAMU_44Ca}, the $K_A$ for the extrema of \nuc{40}{Ca} and \nuc{48}{Ca} follow precisely opposite trends between the two analyses. However, the presence of an additional data point for \nuc{42}{Ca} -- which was absent in the TAMU analysis -- that follows the same general trend as the other three isotopes found in the present work inspires greater confidence in our results. These data, thus, conclusively exclude the possibility of a positive $K_\tau$ value for the calcium nuclei. We also note that the fit presented for the TAMU data in Ref. \cite{TAMU_44Ca} corresponded to $K_{\infty}$ = 200 MeV, which is significantly lower than the currently accepted value of 240$\pm$20 MeV for this quantity~\cite{garg_colo_review, schlomo_knm}.

Also presented in Fig. \ref{RCNP KA} are the $K_A$ values derived from the \nuc{40,42,44,48}{Ca} strength distributions predicted by the FSUGarnet \cite{piekarewicz_private,piekarewicz_calcium_isotopes} relativistic interaction. The $K_\tau^\infty$ has a moderate value of $- 247.3$ MeV for this particular interaction and, accordingly, the $K_A$ values are observed to decrease over the isotopic chain, qualitatively similar to the experimental results. This trend is indeed expected, and observed, for the overwhelming majority of interactions and models \cite{sagawa_private}.

These results, of course, beg the question as to the origin of the differences in the extracted ISGMR responses from those obtained by the TAMU group. The most obvious difference between the experimental techniques lies in the accounting of the instrumental background and physical continuum. Whereas in the present work the former is almost completely eliminated and the physical continuum is included within the extracted ISGMR strength, the TAMU group subtracts both by approximating a smooth background underlying the inelastic spectra. As stated earlier, this has resulted in similar discrepancies in the extracted ISGMR strengths, especially at the higher excitation energies ($E_x > $ 20 MeV), in the $A\sim90$ nuclei \cite{KBH_EPJA,youngblood_A90_unexpected,gupta_A90_PLB,gupta_A90_PRC}.

In summary, motivated by the great cause for concern that would arise were $K_\tau>0$ a reality, we have carried out a systematic measurement of the ISGMR response of \nuc{40,42,44,48}{Ca} and extracted the nuclear incompressibilities, $K_A$, therefrom. In contrast to prior results \cite{TAMU_40Ca,TAMU_44Ca,TAMU_48Ca}, the ISGMR strength distributions, and the metrics that are generally used to characterize the excitation energy of the response, obey expected trends. It may be concluded, therefore, that there are no local structure effects on the ISGMR strength distribution in the calcium region of the nuclear chart and that a {\em positive} value for the asymmetry term of nuclear incompressibility, $K_{\tau}$, is ruled out.

We thank Profs. W. G. Newton, J. Piekarewicz, and H. Sagawa for their comments on the implications of a positive $K_\tau$ in nuclear structure and nuclear astrophysics applications. We are grateful, further, to Prof. J. Piekarewicz for providing the results of the FSUGarnet calculations.
KBH acknowledges the support of the Arthur J. Schmitt Foundation, as well as the Liu Institute for Asia and Asian Studies, and the College of Science, University of Notre Dame. SW would like to thank the Glynn Family Honors program at the University of Notre Dame for financial support. This work has been supported in part by the National Science Foundation (Grant No. PHY-1713857).

\bibliography{UND_calcium}

\end{document}